\documentclass[12pt]{article}

\usepackage{cite}
\usepackage{times}
\usepackage{mathrsfs}
\usepackage{graphicx}
\usepackage{bm}
\usepackage{float}
\usepackage{physics}
\usepackage{siunitx}
\usepackage{xcolor}
\usepackage{subfigure}
\usepackage{appendix}
\usepackage[colorlinks,linkcolor=blue,anchorcolor=blue,urlcolor=blue,citecolor=blue]{hyperref}
\usepackage{multirow}
\usepackage{epstopdf}
\usepackage{amssymb}

\newenvironment{sciabstract}{%
\begin{quote} \bf}
{\end{quote}}

\newcounter{lastnote}

\title{Robust Geometric Gates for QCCD Architecture}

\author
{Xin Zhang$^{1,2}$, Yi Tao$^{1,2}$, Hong Yang Wang$^{1,2}$, Gang Fan$^{1,2}$, \\
Jie Zhang$^{1,2}$, Wei Wu$^{1,2\ast}$, Ting Chen$^{1,2\ast}$, Chun Wang Wu$^{1,2\ast}$, Yi Xie$^{1,2\ast}$,\\
Ping Xing Chen$^{1,2,3}$
\\
\normalsize{$^{1}$College of Science,} \\
\normalsize{National University of Defense Technology, Changsha 410073, Hunan, China}\\
\normalsize{$^{2}$Hunan Key Laboratory of Quantum Information Mechanism and Technology,}\\
 \normalsize{National University of Defense Technology, Changsha 410073, Hunan, China}\\
\normalsize{$^{3}$Hefei National Laboratory, Hefei 230088, Anhui, China}\\
\\
\normalsize{$^\ast$Corresponding authors:}\\
\normalsize{Email: weiwu@nudt.edu.cn(W.W.); chenting17@nudt.edu.cn(T.C.); cwwu@nudt.edu.cn(C.W.);}\\
\normalsize{xieyi2015@nudt.edu.cn(Y.X.)}
}

\date{}

\begin{document}


\baselineskip24pt


\maketitle


\begin{sciabstract}
 The trapped-ion quantum charge-coupled device (QCCD) is a promising route toward scalable quantum computing. However, the frequency offset during shuttling operations introduce significant decoherence and control errors, severely degrading gate fidelity on transported ions. Here, using a room-temperature planar chip trap, we demonstrate a robust and universal single-qubit gate set in the QCCD architecture. Based on a recently proposed path-optimizable noncyclic geometric scheme, we show that dynamic noise is strongly suppressed even when multiple shuttling operations are interleaved with gate execution, achieving substantially higher fidelities than conventional dynamical gates. This robustness is further validated in a multi-ion system containing up to 33 qubits. With inherent robustness and adaptability to varying shuttle distances and ion numbers, our work offers a plug-and-play solution for building robust quantum primitives for the QCCD platform.
\end{sciabstract}


The quantum charge-coupled device (QCCD) architecture, composed of functionally distinct zones hosting shuttled ions as qubits, is among the most promising routes for large-scale quantum computation~\cite{ArchitectureLIKD,CompletemsHJP,HighfidelitytBRB,Benchmarking11WK,DemonstrationtiPJM,FaulttoletantPRHJ,RacetracktiMSA,A98qubittiRA}. However, a fundamental bottleneck persists: ions shuttled between zones inevitably experience complex, spatiotemporally varying noise environments—such as unwanted inhomogeneous magnetic fields~\cite{SingleisWA,FastISKH}—leading to gate errors~\cite{ExperimentalicWDJ,LonglivedztiRT}. Existing mitigation strategies, including magnetic-field geometry optimization for improved homogeneity~\cite{DevelopmentHPJFW}, zone-by-zone frequency recalibration~\cite{SingleisWA}, and dynamical decoupling~\cite{SpinechoHEL,RobustCJ}, incur substantial experimental overheads and provide only limited protection, leaving transporting qubits intrinsically more fragile than their stationary counterparts.

Geometric quantum gates~\cite{AppearanceGSWF,QuantalPFBMV,HolonomyQATSB,PhaseAA,HolonomcQCPZ,GeometricMTDLM,ExperimentalDRLD,GeometricPGKK,NonadiabaticHQSE,NonadiabaticCGXBW,ExperimentalRNFG,ExperimentalRNAJAA,DoublyGQDWZ,GeometricHQZJ,RecentANXZY}, constructed from the global geometric properties of the evolution path, are inherently insensitive to local dynamical details, offering a natural route toward robust quantum operations. While early demonstrations~\cite{GeometricQCJJA,ExperimentalDBMM,ExperimentalDSFS} relied on adiabatic and cyclic evolution, subsequent developments of nonadiabatic noncyclic geometric gates~\cite{NoncyclicGQFA,ErrortolerantCT,GenuinelyNGEN,SinglepulseTGCZY} have removed these constraints while preserving geometric robustness, and have been realized in superconducting qubits~\cite{NoncyclicNGMZ}, and trapped ions~\cite{SingleatomVNZJW}. Existing implementations typically depend on a single, fixed evolution path for a given gate, leaving no degrees of freedom for optimization against different noise sources. Recently, a path-optimizable geometric gate scheme was proposed~\cite{RobustnessEUQZH}, which provides a family of candidate evolution paths for the same gate, allowing one to select the optimal path tailored to the prevailing noise environment and thereby enhance the robustness of universal noncyclic geometric gates. Whether this advantage can be translated into high-fidelity quantum logic gates in the QCCD architecture, which features a far more complex noise environment than stationary qubits, constitutes a strong experimental motivation.

Here, in a shuttling-based QCCD setting using a room-temperature surface-electrode trap, we realize a universal single-qubit gate set---$R_x$ and $R_y$ rotations---based on the path-optimizable noncyclic geometric scheme, and demonstrate a significant robustness advantage over conventional dynamical gates. Using a controllable magnetic gradient as a tunable noise source, we benchmark the optimized geometric gates against dynamical ones as ions shuttle across zones, observing pronounced error suppression without recalibration. Extending these gates to global operations on multi-ion crystals of up to 33 ions confirms that the robustness advantage persists and grows with system size. These results provide the first experimental demonstration of path-optimizable geometric gates and highlight their strong potential for scalable QCCD quantum computing.

\section*{RESULTS}
\subsection*{Path-optimizable nonadiabatic noncyclic geometric gates}
Consider a generic two-level system with basis 
states $\{\ket{0},\ket{1}\}$ separated by an energy splitting $\hbar\omega_0$, driven by a coherent field with frequency $\omega_d(t)$ and phase $\phi(t)$. In the frame rotating at $\omega_d(t)$, and under the 
rotating-wave approximation, the Hamiltonian reads
\begin{equation}
\mathcal{H}(t)=\frac{1}{2}
\begin{pmatrix}
-\Delta(t) & \Omega(t)\,e^{-i\phi(t)} \\[1mm]
\Omega(t)\,e^{i\phi(t)} & \Delta(t)
\end{pmatrix},
\label{eq:Hamiltonian}
\end{equation}
where $\Delta(t)=\omega_0-\omega_d(t)$ is the detuning and $\Omega(t)$ is the Rabi frequency. Following the
formalism of Refs.~\cite{NonadiabaticHQSE,NoncyclicGQFA,RobustnessEUQZH}, we introduce a pair of
orthogonal auxiliary states parametrized by coordinates
$(\kappa,\xi)$ on the Bloch sphere:
\begin{subequations}
\begin{align}
\ket{\Psi_{1}(t)}
&= e^{if_{1}(t)}[\cos\frac{\kappa(t)}{2}\,\ket{0}
+ \sin\frac{\kappa(t)}{2}\,e^{i\xi(t)}\,\ket{1}],
\label{eq:psi1} \\[2mm]
\ket{\Psi_{2}(t)}
&= e^{if_{2}(t)}[\sin\frac{\kappa(t)}{2}\,e^{-i\xi(t)}\,\ket{0}
- \cos\frac{\kappa(t)}{2}\,\ket{1}].
\label{eq:psi2}
\end{align}
\label{eq:aux-states}
\end{subequations}
Requiring $\ket{\Psi_{m}(t)} (m=1,2)$ to satisfy the Schr\"odinger equation governed by $\mathcal{H}(t)$ yields 
\begin{align}
\dot{\kappa} &= \Omega(t)\sin[\phi(t)-\xi(t)],
\label{eq:kappa-dot}\\
\dot{\xi}   &= -\Delta(t)
  - \Omega(t)\cot\kappa(t)\cos[\phi(t)-\xi(t)],
\label{eq:xi-dot}
\end{align}
and during the evolution period $\tau$ the total accumulated phase 
$\gamma = f_{1}(\tau) = -f_{2}(\tau) = \frac{1}{2}\!\int_0^\tau\!\frac{\dot{\xi}(t)[1-\cos\kappa(t)] + \Delta(t)}{\cos\kappa(t)}\,dt$, comprising a dynamical part
$\gamma_d = \frac{1}{2}\!\int_0^\tau\!
\frac{\dot{\xi}(t)\sin^2\!\kappa(t) + \Delta(t)}{\cos\kappa(t)}\,dt$ 
and a geometric part
$\gamma_g = -\frac{1}{2}\int_0^\tau
\dot{\xi}(t)\,(1-\cos\kappa(t))\,dt
= \frac{1}{2}\,\Omega_{\mathrm{sol}}$,
where $\Omega_{\mathrm{sol}}$ is the solid angle enclosed by
the evolution trajectory together with the geodesic connecting its endpoint and starting point on the Bloch sphere~\cite{PhaseAA}. 
The time-evolution operator then decomposes as
\begin{equation}
U(\tau) = \ket{\Psi_1(\tau)}\!\bra{\Psi_1(0)}
        + \ket{\Psi_2(\tau)}\!\bra{\Psi_2(0)}.
\label{eq:U-aux}
\end{equation}
By enforcing $\gamma_d=0$, the evolution operator $U(\tau)$ becomes governed purely
by $\gamma_g$, which depends only on the trajectory geometry
$\{\kappa(t),\xi(t)\}$ and does not explicitly contain $\Delta$. Although a detuning offset $\delta\Delta$ perturbs the actual trajectory via Eq.~\eqref{eq:xi-dot}, the absence of
$\Delta$ from $\gamma_g$ strongly suppresses the gate's
sensitivity to detuning errors compared with its dynamical
counterpart~\cite{GeometricHQZJ,RecentANXZY}.

\begin{figure}[tbhp]
    \centering
    \includegraphics[width=0.85\columnwidth,trim=0cm 8.3cm 0cm 0.5cm,clip]{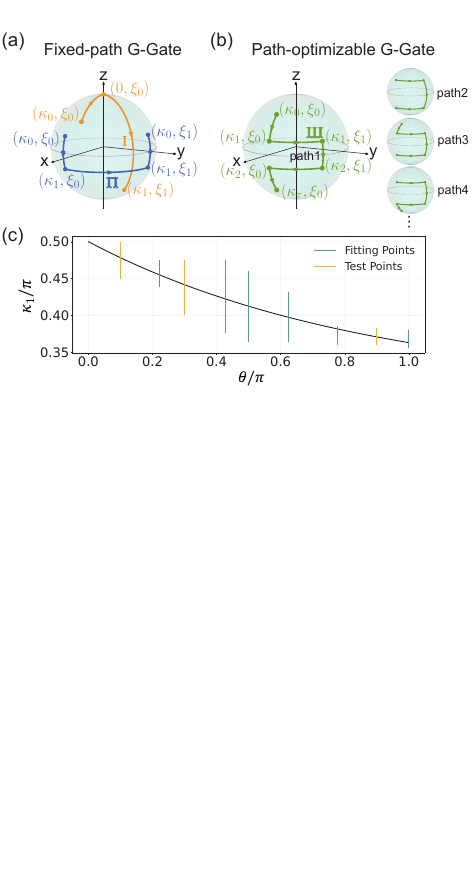}
    \caption{(a) Fixed-path geometric gates (G-gates) [cases~\hyperref[eq:case-I]{I} and \hyperref[eq:case-II]{II} in Eq.~\eqref{eq:noncyclic-cases}]: a given gate corresponds to a unique evolution path. 
    (b) Path-optimizable G-gate: for case \hyperref[eq:case-III]{III}, multiple evolution paths can be constructed, and path~1 can realize the desired $R_x(\theta)$ and $R_y(\theta)$ gates while retaining free parameters $\kappa_1$ and $\kappa_2$ for robustness optimization. (c) Under the constraint $\kappa_2=\pi-\kappa_1$, the high-fidelity intervals of $\kappa_1$ (where $\overline{F}(\kappa_1)\ge 0.998\,\overline{F}_{\max}$, with $\overline{F}(\kappa_1)$ and $\overline{F}_{\max}$ the $\kappa_1$-dependent fidelity and the maximal fidelity under a generic error model, respectively) are numerically obtained for $\theta=2\pi/9,\,3\pi/7,\,\pi/2,\,5\pi/8,\,7\pi/9,\,\pi$ (green bars); an exponential fit to their midpoints yields the empirical curve. The curve also falls within the high-fidelity intervals for three arbitrarily chosen additional angles $\theta=0.1\pi,\,0.3\pi,\,0.9\pi$ (orange bars, not included in the fit), validating the practicality of this empirical formula.}
    \label{fig1}
\end{figure}

Following Ref.~\cite{RobustnessEUQZH}, noncyclic evolution paths can be classified into three types according to the relative positions of their initial and final points on the Bloch sphere:
\begin{subequations}
	\label{eq:noncyclic-cases}
	\begin{align}
		\text{(I)}\quad
		& \kappa(\tau) \neq \kappa(0), \qquad
		\xi(\tau) \neq \xi(0) + 2n\pi,
		\label{eq:case-I}\\
		\text{(II)}\quad
		& \kappa(\tau) = \kappa(0), \qquad
		\xi(\tau) \neq \xi(0) + 2n\pi,
		\label{eq:case-II}\\
		\text{(III)}\quad
		& \kappa(\tau) \neq \kappa(0), \qquad
		\xi(\tau) = \xi(0) + 2n\pi,
		\label{eq:case-III}
	\end{align}
\end{subequations}
where $n\in\mathbb{Z}$. For simplified piecewise trajectories composed of meridional and latitudinal segments, cases~\hyperref[eq:case-I]{I} and \hyperref[eq:case-II]{II} yield a unique evolution path determined by the target gate, leaving no free parameters for robustness optimization [Fig.~\hyperref[fig1]{\ref*{fig1}(a)}]. For case~\hyperref[eq:case-III]{III}, multiple trajectories exist [Fig.~\hyperref[fig1]{\ref*{fig1}(b)}], among which path~1 can directly realize noncyclic $R_x(\theta)$ and $R_y(\theta)$ gates while retaining enough free parameters for optimization; we therefore adopt it to construct our geometric gates. With $\kappa(\tau)-\kappa(0)=\theta$, $\gamma_g=\pi$, and $\xi(\tau)=\xi(0)\equiv\xi_0$, the choices $\xi_0=-\pi/2$ and $\xi_0=0$ realize $R_x(\theta)$ and $R_y(\theta)$, respectively. We impose the constraint $\kappa_0+\kappa_\tau=\pi$, empirically found to yield higher gate robustness  and use a constant Rabi frequency $\Omega$ throughout the evolution for experimental simplicity. For a given $\theta$, the candidate path is then parameterized by two intermediate polar angles, $\kappa_1$ and $\kappa_2$. 

To identify the most robust path, we numerically optimize $\kappa_1$ and $\kappa_2$ by maximizing the average gate fidelity under a generic static error Hamiltonian $\hat H_{\mathrm{err}} = \lambda_m \, \vec{n} \cdot \hat{\vec{\sigma}}$~\cite{RobustQGYK}, where $\vec{n}$ is a randomly sampled unit vector and $\lambda_m$ controls the error strength, set to $0.05\,\Omega$ in this work. The fidelity is averaged over 100 independent random samples, and the path yielding the highest average fidelity is deemed the most robust. We find that the optimized paths generally satisfy $\kappa_1 + \kappa_2 \approx \pi$. Motivated by this observation, we impose the additional constraint $\kappa_2 = \pi - \kappa_1$ in the subsequent optimization, thereby reducing the candidate path to a single free parameter $\kappa_1$.

While Ref.~\cite{RobustnessEUQZH} considered robustness optimization only for a fixed $\pi$ rotation, we extend the optimization to a range of rotation angles and establish an empirical prescription for selecting near-optimal paths at arbitrary angles. For rotation angles $\theta = \{2\pi/9,\allowbreak\; 3\pi/7,\allowbreak\; \pi/2,\allowbreak\; 5\pi/8,\allowbreak\; 7\pi/9,\allowbreak\; \pi\}$, we sweep the single free parameter $\kappa_1$ and identify the high-fidelity intervals where $\overline{F}(\kappa_1) \ge 0.998\,\overline{F}_{\max}$, with $\overline{F}$ the $\kappa_1$-dependent average gate fidelity and $\overline{F}_{\max}$ the maximum fidelity obtained from the $\kappa_1$ optimization. Taking the midpoints of these intervals as fitting points, an exponential fit yields the empirical relation (in units of $\pi$)
\begin{equation}
	\kappa_1^{\mathrm{fit}}(\theta)=0.201063\,e^{-1.142991\,\theta}+0.298937,
	\label{eq:kappa_fit}
\end{equation}
as shown in Fig.~\hyperref[fig1]{\ref*{fig1}(c)}. To verify the practicality of this empirical formula, we arbitrarily choose three additional angles $\theta = \{0.1\pi,\, 0.3\pi,\, 0.9\pi\}$ and compute their high-fidelity intervals; the fitted curve falls within all these intervals. This validates the empirical formula as a practical prescription for rapidly selecting a near-optimal noncyclic geometric path for any desired rotation angle $\theta$ via a closed-form expression.

\subsection*{Experimental setup and gate benchmarking on static qubits}
\begin{figure*}[htbp]
	\centering
	\includegraphics[width=\textwidth,trim=0cm 4.3cm 0cm 0.55cm,clip]{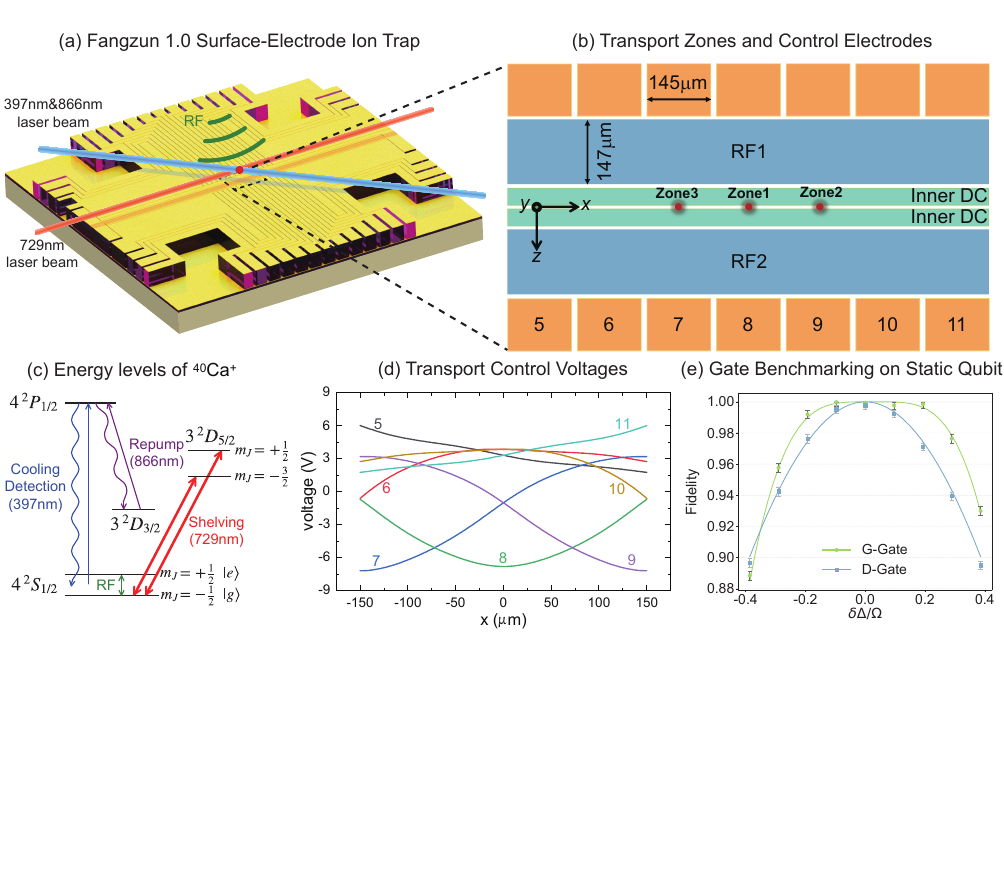}
	\caption{(a) Schematic of the Fangzun 1.0 surface-electrode ion trap. (b) Spatial distribution of the transport zones and primary control electrode pairs 5–11. (c) Energy levels of $^{40}\mathrm{Ca}^+$ and the relevant driving lasers and RF fields. (d) Voltage settings of control electrode pairs 5–11 versus ion equilibrium position. Ion transport is enabled by applying time-dependent voltage settings. (e) Fidelity comparison of geometric(G-) and dynamical (D-)gates on a static qubit as a function of detuning offset.}
	\label{fig2}
\end{figure*}

Our experiments are performed on a room-temperature surface-electrode ion trap, Fangzun 1.0 [Fig.~\hyperref[fig2]{\ref*{fig2}(a,b)}]~\cite{CoherentMMXY}. The chip features a six-wire geometry comprising a pair of RF electrodes, a pair of inner DC electrodes, and $17$ pairs of outer DC control electrodes. Ions are trapped $\sim$150~$\mu$m above the surface and illuminated by the required lasers and an RF field generated by a coil 2~cm above the chip. Pairs 5--11 of the outer DC electrodes are applied time-dependent voltages for shuttling ions across zones [Fig.~\hyperref[fig2]{\ref*{fig2}(d)}]; the rest pairs are used to compensate stray fields. The qubit is encoded in the Zeeman sublevels $4^{2}S_{1/2}$ ($m_J=\pm1/2$) of $^{40}\mathrm{Ca}^+$, driven by a near-resonant RF field [Fig.~\hyperref[fig2]{\ref*{fig2}(c)}]. State readout uses a two-step electronic shelving sequence and the subsequent fluorescence detection. First, a 729~nm laser transfers population from $4^{2}S_{1/2}(m_J=-1/2)$ to $3^{2}D_{5/2}(m_J=-3/2)$; a second 729~nm laser then transfers the residual population to $3^{2}D_{5/2}(m_J=+1/2)$. After shelving, fluorescence on the 397~nm transition is collected to determine the $4^{2}S_{1/2}(m_J=+1/2)$ population. An 866~nm laser repumps the $D_{3/2}$ population back to the $S_{1/2}$ manifold. An arbitrary waveform generator (AWG) independently controls the strength, detuning, and phase of both the RF and 729~nm laser fields. All qubit gates and shelving operations ($\pi$ rotations) are implemented using the five-segment noncyclic evolution path shown in Fig.~\hyperref[fig1]{\ref*{fig1}(b)}, and the independently characterized shelving errors have been subtracted from all gate fidelities reported below(see Supplementary). The optimal $\kappa_1$ for each rotation angle $\theta$ is chosen following the robust optimization procedure in the previous section. The Rabi frequency $\Omega$ is fixed at $2\pi\times20.7\ \mathrm{kHz}$, and the evolution path is controlled solely through $\Delta(t)$ and $\phi(t)$, as determined by Eqs.~\eqref{eq:kappa-dot} and~\eqref{eq:xi-dot}(see Supplementary).

We first characterize four geometric gates $R_x(3\pi/7)$, $R_x(7\pi/9)$, $R_y(2\pi/9)$, and $R_y(5\pi/8)$ on a stationary ion via quantum process tomography (QPT)~\cite{PrescriptionEDILC}, with $10\,000$ repetitions per data point. The maximum-likelihood process fidelities~\cite{MeasurementQJDFV}(see Supplementary) are $1.0000_{-0.0025}^{+0.0000}$, $1.0000_{-0.0024}^{+0.0000}$, $1.0000_{-0.0023}^{+0.0000}$, and $0.9995_{-0.0023}^{+0.0005}$, respectively, demonstrating near-unity fidelities for our geometric gates without detuning errors.

In the shuttling scenario, the dominant gate error is due to position-dependent detuning offsets. Fig.~\hyperref[fig2]{\ref*{fig2}(e)} compares the fidelity of geometric and dynamical $R_y(5\pi/8)$ gates on a stationary qubit versus detuning offset. The geometric gate shows significantly enhanced robustness against detuning errors than its dynamical counterpart. At zero detuning, they show comparable fidelities of $0.9995_{-0.0023}^{+0.0005}$ and $0.9978_{-0.0025}^{+0.0022}$, respectively. However, the geometric gate remains above $0.98$ throughout $-0.23 < \delta\Delta/\Omega < 0.28$ and exceeds the dynamical gate across the full measured range $|\delta\Delta|/\Omega \le 0.35$, which already covers the offsets expected in the shuttling and multi-ion experiments.

\subsection*{Geometric gates on transporting qubits in QCCD architecture}
\begin{figure*}[htbp]
	\centering
	\includegraphics[width=\textwidth,trim=0cm 5.0cm 0cm 0.85cm,clip]{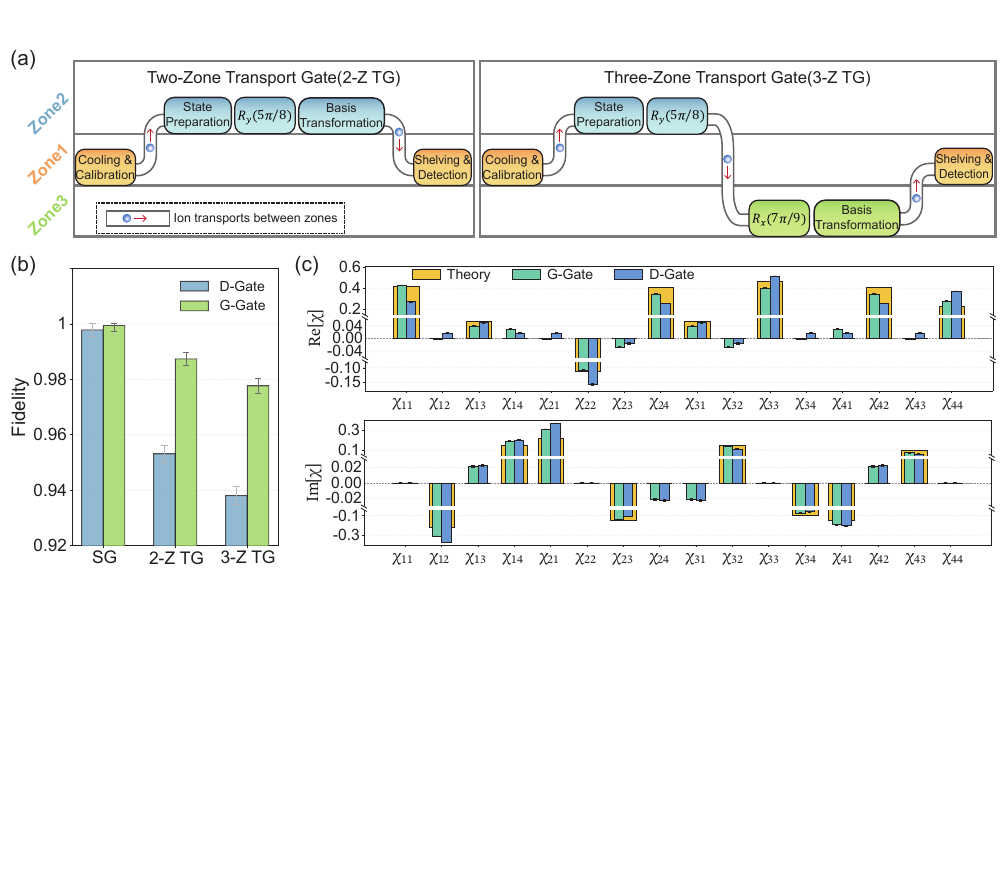}
	\caption{(a) Procedures for the two-zone transport gate (2-Z TG) and three-zone transport gate (3-Z TG). (b) Fidelity comparison of geometric and dynamical gates for the stationary gate (SG), 2-Z TG, and 3-Z TG. (c) Real and imaginary parts of the reconstructed process matrices $\chi$ for the geometric and dynamical 3-Z TG, compared with the ideal theoretical values.}
	\label{fig3}
\end{figure*}

To assess the robustness of our geometric gates in the QCCD architecture, we introduce a magnetic-field gradient of $\sim\!11.7\;\mathrm{Gauss/m}$ along the trap axis, leading to severe frequency offset errors during shuttling. Zone~1 is the loading and cooling region; zones~2 and~3 ($\pm150\;\mu$m displacement, and ${\sim}\pm5.25\;\mathrm{kHz}$ offset,respectively) are operation regions [Fig.~\hyperref[fig2]{\ref*{fig2}(b)}]. We compare the performance of geometric and dynamical gates for two transport scenarios against the stationary scenario. For the stationary gate (SG), cooling, calibration and an $R_y(5\pi/8)$ gate are performed on a stationary qubit in zone~1. For the two-zone transport gate (2-Z TG) [Fig.~\hyperref[fig3]{\ref*{fig3}(a)}, left], cooling and calibration occur in zone~1; the qubit shuttles to zone~2 for state preparation, $R_y(5\pi/8)$ gate, and basis transformation, then returns to zone~1 for shelving and detection. For the three-zone transport gate (3-Z TG) [Fig.~\hyperref[fig3]{\ref*{fig3}(a)}, right], after cooling and calibration in zone~1, the qubit moves to zone~2 for state preparation and $R_y(5\pi/8)$, then to zone~3 for $R_x(7\pi/9)$ and basis transformation, and finally back to zone~1 for shelving and detection.

Fidelity results are shown in Fig.~\hyperref[fig3]{\ref*{fig3}(b)}. For SG, the geometric and dynamical fidelities are $0.9995_{-0.0023}^{+0.0005}$ and $0.9978_{-0.0025}^{+0.0022}$, respectively. For 2-Z TG, fidelities drop to $0.9874(24)$ (geometric) and $0.9531(30)$ (dynamical); the dynamical gate exhibits $3.7$ times the infidelity of the geometric gate. For 3-Z TG, fidelities are $0.9777(27)$ (geometric) and $0.9381(32)$ (dynamical), yielding an infidelity ratio of $2.8$. The reconstructed process-matrix elements for the geometric and dynamical 3-Z TG are compared with the ideal values in Fig.~\hyperref[fig3]{\ref*{fig3}(c)}. Most of sizable matrix elements lie closer to the ideal values for the geometric gate than for the dynamical gate, demonstrating a clear fidelity advantage and confirming the practical benefit of geometric gates in shuttling scenarios.

\subsection*{Geometric gates on multi-ion crystals}
A single potential well on the chip can simultaneously confine multiple ions. When driven globally, the magnetic-field gradient introduces a position-dependent detuning offset on each ion, degrading the overall gate fidelity. We examine whether globally driven geometric gates maintain high fidelity across different crystal configurations. Fluorescence is averaged over all ions to reconstruct an ensemble density matrix via quantum state tomography(see Supplementary). The gate fidelity is quantified by the state fidelity between the ideal target state $R_y(5\pi/8)\ket{g}$ and the experimentally reconstructed density matrix. We benchmark these gates against their dynamical counterparts under identical conditions.

\begin{figure}[htbp]
	\centering
	\includegraphics[width=0.85\columnwidth,trim=0cm 11cm 0cm 0cm,clip]{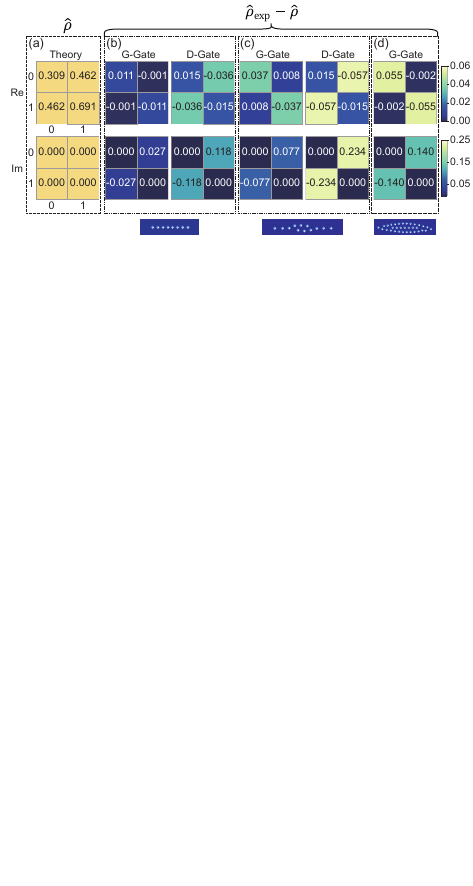}
	\caption{(a) Ideal density matrix of the target state $\hat{\rho}$. (b)--(d) Deviation of the experimentally reconstructed density matrix $\hat{\rho}_{\text{exp}}$ from the ideal target for geometric and dynamical gates on the 8-ion linear chain, 12-ion zigzag, and 33-ion planar crystal, respectively.}
	\label{fig4}
\end{figure}

The results are shown in Fig.~\ref{fig4}. For an 8-ion linear chain (detuning offset spread $\sim\pm1\;\mathrm{kHz}$), the average state fidelities are $0.9947(25)$ (geometric) and $0.9611(31)$ (dynamical), with the dynamical gate exhibiting $7.3$ times the infidelity. For a 12-ion zigzag crystal (detuning offset spread $\sim\pm1.2\;\mathrm{kHz}$), the fidelities are $0.9927(08)$ (geometric) and $0.9419(23)$ (dynamical), yielding an infidelity ratio of $8.0$. Given this systematic advantage at the 8- and 12-ion scales, and the challenge of stably trapping 33 ions over long acquisition periods due to occasional collision-induced losses, we measured only the geometric gate at this scale. For the 33-ion planar crystal, we obtain a state fidelity of $0.9775(14)$, confirming that the geometric gate maintains good operational quality at a practically relevant crystal size.
  
\section*{DISCUSSION}
This work presents the first integration of geometric approach with the QCCD architecture, a promising platform for scalable quantum computing. Using the path-optimizable nonadiabatic noncyclic geometric scheme, we demonstrate a robust and universal single-qubit gate set in the shuttling-based operation setting. Under realistic noise conditions, the geometric gate consistently accumulates several-fold less error than its dynamical counterpart. In multi-ion crystals up to 33 ions, the geometric gate maintains high fidelity, with its advantage growing with crystal size. These robust primitives can be directly applied to various QCCD operations, including state preparation, basis rotation, gate benchmarking, etc. A natural next step is to extend the geometric approach to two-qubit entangling gates, where shuttling-induced frequency offsets pose an analogous challenge. Combining detuning-resilient single- and two-qubit gates would yield a robust, complete gate set for the QCCD architecture.

\section*{Acknowledgments}
\textbf{Funding:} This work was supported by the National Natural Science Foundation of China (Grants No. 12174447, No. 12174448, No. 12204543 and No. 11904402), Quantum Science and Technology-National Science and Technology Major Project (QNMP) (Grant No.2021ZD0301605), the Science and Technology Innovation Program of Hunan under Grant No. 2022RC1194, and the Natural Science Foundation of Hunan Province under Grant No.2023JJ10052. 

\noindent{\textbf{Author contributions:}} W.W. and P.C. supervised the whole project. C.W., Y.X., T.C. and X.Z. contributed to the conception of the work, the analysis and interpretation in the work, and the preparation and revision of the manuscript. J.Z., Y.T., H.W. and G.F. contributed to building the experimental setup.

\noindent{\textbf{Competing interests:}} The authors declare no competing financial interests.

\noindent{\textbf{Data availability:}} Data that support the plots within this paper are available from the corresponding authors upon reasonable request. Source data are provided with this paper.

\noindent{\textbf{Code availability:}} Code that related to this paper are available from the corresponding authors upon reasonable request.

\bibliography{article_management} 

\clearpage

\newpage

\newpage

\newpage

\end{document}